\documentclass[aps,times,prb,a4paper,twocolumn,amsmath,amssymb,floatfix,showpacs]{revtex4}

\usepackage[dvips]{graphicx}

\begin{document}

\title{Density of quasiparticle states for a two-dimensional disordered
system:\\ Metallic, insulating, and critical behavior 
in the class-D thermal quantum Hall effect} 
\author{A. Mildenberger,$^{1,2}$ F. Evers,$^{2,3}$ 
A.~D. Mirlin,$^{2,3,*}$ and  J.~T. Chalker$^{3,4}$} 
\affiliation{
$\mbox{}^{1}$Fakult\"at f\"ur Physik,~Universit\"at Karlsruhe,~76128
Karlsruhe,~Germany\\ 
$\mbox{}^{2}$Institut f\"ur Nanotechnologie, Forschungszentrum
Karlsruhe, 76021 Karlsruhe, Germany \\ 
$\mbox{}^{3}$Institut f\"ur Theorie der Kondensierten Materie,
Universit\"at Karlsruhe, 76128 Karlsruhe, Germany\\
$\mbox{}^{4}$Theoretical  Physics, University 
of Oxford, 1 Keble Road, Oxford OX1 3NP, UK }
\date{\today}

\begin{abstract}
We investigate numerically the quasiparticle density of states
$\varrho(E)$ for a two-dimensional, disordered superconductor in which
both time-reversal and spin-rotation symmetry are broken. As a generic
single-particle
description
of this class of systems (symmetry class D), we use the Cho-Fisher version of
the network model. This has three phases: a thermal insulator, a
thermal metal, and a quantized thermal Hall conductor.
In the thermal metal, we find a logarithmic divergence in
$\varrho(E)$ as $E\to 0$,
as predicted from sigma model calculations. Finite-size effects lead
to superimposed 
oscillations, as expected from random matrix theory.
In the thermal insulator and quantized thermal Hall conductor, we find
that $\varrho(E)$ 
is finite at $E=0$. At the plateau transition 
between these phases, $\varrho(E)$ decreases toward zero
as $|E|$ is reduced, in line with the result $\varrho(E) \sim |E|\ln(1/
|E|)$ derived from calculations for
Dirac fermions with random mass. 
\end{abstract}

\maketitle
%\narrowtext

%%%%%%%%%%%%%%%%%%%%%%%%%%%%%%%%%%%%%%%%%%%%%%%%%%%%%%%%%%%%%%%%
%%%%%%%%% Introduction
%%%%%%%%%%%%%%%%%%%%%%%%%%%%%%%%%%%%%%%%%%%%%%%%%%%%%%%%%%%%%%%%
\section{Introduction}
\label{s1}

The study of Anderson transitions -- 
disorder-driven transitions in systems of noninteracting fermions --
has a long history in condensed-matter physics.
Symmetry is a guide to understanding both mesoscopic 
behavior in each phase and critical behavior at a transition. 
According to the symmetry classification inherited from random-matrix
theory, in a conventional situation three Wigner-Dyson symmetry classes --
orthogonal, unitary and symplectic --
can be distinguished, 
depending on whether or not the Hamiltonian is invariant under
time reversal and spin rotation (see Refs.
\onlinecite{lee,kramer,guhr,efetov,adm-physrep} for reviews). 
Above a lower critical dimension, systems in each symmetry class
can have a transition between insulating and metallic phases.
In addition, in two-dimensional systems without symmetry under
time-reversal or parity, a second type of insulating phase
is possible, with edge states and a quantized Hall conductance,
and there can be a plateau transition between the insulator
and the quantized Hall conductor.
Long-distance properties of these systems are described by
nonlinear sigma models. Within this
framework, quantized Hall conductors appear if
symmetry allows a topological term in the sigma model.\cite{huckestein95}

All Anderson transitions in the
standard, Wigner-Dyson symmetry classes are expected to share 
certain general features. One is that the density of states (DoS)
is a smooth function of energy, and hence
noncritical at a transition. Another is universality, in the
sense that, for a given symmetry and dimensionality, only one type of
behavior is expected.

It is now widely appreciated that the range of possibilities is
not exhausted by the Wigner-Dyson classes,
and there exist additional symmetry
classes. These additional classes are  distinguished from the
standard ones by possessing a particle-hole symmetry that selects
as special one particular energy $E$ in the spectrum (we take this
to be $E=0$). Moreover, their density of states may be singular at
$E=0$ and may have distinct behavior in each phase and at a critical
point. While individual examples of systems from the additional symmetry
classes have been known for many years, as tight-binding models with two
sublattice structure,\cite{dyson,eggarter,gade}
within random-matrix theory,\cite{slevin,verbaarschot}
and as models for superconductors,\cite{oppermann}
a full classification scheme was developed only more recently
by Altland and Zirnbauer.
\cite{altland,zirnbauer} 
This scheme includes, in addition to the Wigner-Dyson classes, systems
belonging to two
additional types of symmetry class. One set, the chiral classes, arises
in two sublattice systems. The other set is realized in models
for noninteracting quasiparticles in disordered superconductors based on
Bogoliubov-de Gennes Hamiltonians.
The properties of models belonging to these additional symmetry
classes have been a focus of attention in
connection with understanding low-energy quasiparticle DoS, transport,
and localization 
properties in dirty superconductors, including those with unconventional 
pairing.\cite{altland,sentil99,senthil99a,
motrunich00,motrunich01,hirschfeld,%%
altland00,bocquet00,kagalovsky,chalker00}  
Among other features, the additional symmetry classes
allow for versions of the quantum Hall effect (QHE) 
in two-dimensional disordered superconductors with broken
time-reversal symmetry. The relevant quantized conductance in
these cases is a thermal or spin conductance, since quasiparticle number is not
conserved by the Hamiltonian.
For superconductors that are invariant
under spin-rotations, the symmetry is referred to as class C,  
and the version of the quantum Hall effect is known as the spin QHE.
Without spin rotation invariance, the symmetry 
is termed  class D and one has the thermal QHE.
The Hamiltonian for a system of the class D has the following block
structure in the particle-hole space \cite{altland}:
\begin{equation}
\label{e0}
H= \left(\begin{array}{cc} h          & \Delta 
                           \\-\Delta^* & - h^T
          \end{array} \right)\ , \ \ 
h = h^\dagger\ , \ \ \Delta= - \Delta^T\ ,
\end{equation}
which is determined by the condition $H=-\sigma_x H^T \sigma_x$ (in
addition to the Hermiticity $H=H^\dagger$).  Alternatively, one can work in
a different basis, defining $\tilde{H}= g^\dagger H g$ with $g^2 = \sigma_x$. 
In this basis, the defining condition of class D becomes $\tilde{H} = - \tilde{H}^T$,
so that $\tilde{H}$ is purely imaginary. It is the systems with this symmetry that
are the subject of the present paper.

A striking feature of this class is that 
symmetry and dimensionality alone are insufficient to 
determine behavior. At the level of the nonlinear sigma model,
the reason is believed to be that the relevant target space has 
two disconnected pieces and that, depending on the choice of underlying
microscopic model, it may or may not be necessary to
consider configurations containing
domain walls on which the sigma model field jumps between the two 
components.\cite{bocquet00,read,gruzberg}
In the following, we investigate the
Cho-Fisher (CF) network model\cite{cho97} for the thermal QHE.
This model is generic in the sense that it displays all
three phases possible in a two-dimensional
class D system: the metal, 
the insulator, and the quantized Hall conductor.
By contrast, in a closely related model -- a
fermionic version of the $\pm J$ random bond Ising model (RBIM) --
the metallic phase is absent.\cite{read}

\begin{figure}[tb]
  \includegraphics[width=8cm]{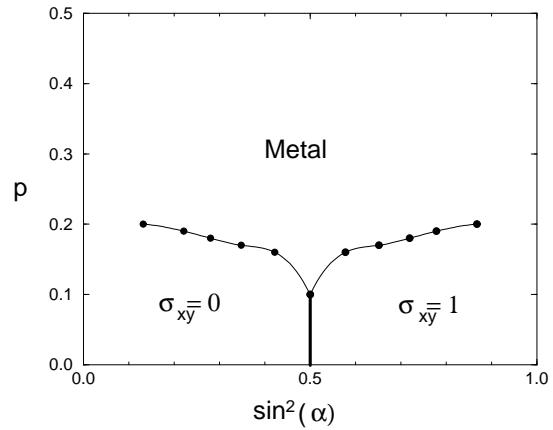}
\caption{Phase diagram of the Cho-Fisher model as obtained in
  Ref.~\onlinecite{chalker00} from transfer-matrix calculations.
  The plane is spanned by the parameters $\sin^2(\alpha)$, the
  interplaquette
tunneling probability, 
and $p$, the concentration of vortex disorder: these control the
short-distance values of the
conductivity components $\sigma_{xy}$ and  
$\sigma_{\rm  xx}$, respectively.}
\label{f1}
\end{figure}

The location of the metallic phase and the two localized phases
in the phase diagram of the CF model is
indicated in Fig.~\ref{f1},
together with phase boundaries
of two types, corresponding to metal-insulator and
plateau transitions, respectively. This phase diagram was obtained in
Ref.~\onlinecite{chalker00} from transfer-matrix calculations; let us
explain why its structure may be quite naturally anticipated. In the
absence of disorder ($p=0$) the system has a gap around $E=0$ for all
$\alpha$ except for $\sin^2\alpha={1\over 2}$ (where the DoS vanishes
linearly at zero energy). One thus expects that the system remains
insulating for weak disorder (small $p$). More carefully, analyzing
the edge states in the both limits of uncoupled plaquettes,
$\sin^2\alpha=0$ and $\sin^2\alpha=1$, one observes that there are two
topologically different insulating phases. From  symmetry, if there is a direct transition
between these phases, it must occur at $\sin^2\alpha={1\over
  2}$. Finally, with increasing $p$, the disorder creates a substantial
DoS at $E=0$, and the system may be expected to undergo a transition
into the metallic phase, which generically exists in two-dimensional
 systems with spin-orbit coupling. 

%%%%%%%%%%%%%%%%%%%%%%%%%%%%%%%%%%%%%%%%%%%%%%%%%%%%%%%%%%%%%%%%
%%%%%%%%%%%%%%%%%%%%%%%%%%%%%%%%%%%%%%%%%%%%%%%%%%%%%%%%%%%%%%%%

Two-dimensional systems from class D have attracted considerable
attention recently. The sigma model analysis
\cite{senthil99a,bocquet00} 
starts from a short-distance description in terms of
a diffusive metal with a smooth DoS that is finite at $E=0$.
From a renormalization group (RG) analysis
(perturbative in $g^{-1}$, the inverse dimensionless conductance),
it is found that the diffusion constant is unrenormalized at leading
order and that 
the DoS has a logarithmic divergence near zero energy, which is
given in terms of the diffusion constant $D$ and the mean free path
$\ell_0$
by 
\begin{equation}
 \varrho(E) = \varrho_0 + \frac{1}{4\pi^2D}\ln
 \frac{D}{|E| \ell_0^2}\,.
%
%\varrho(E) = \frac{1}{8\pi^2D}\ln
% \left( \frac{D}{2E\ell_0^2} \right)^2 + \mbox{regular}
%
\label{e1}
\end{equation}

An alternative approach to the theory of these systems, arguably
tailored to describe the plateau transition, has been developed in
Ref.~\onlinecite{bocquet00} by starting from a model of Dirac fermions
with random mass and treating this disorder perturbatively,
in the spirit of the analysis of the Ising model by
Dotsenko and Dotsenko.\cite{dotsenko83} 
The disorder-free system has a transition, driven by tuning 
a uniform mass through zero, which in the CF model
lies at $p{=}0$, $\sin^2(\alpha){=}1/2$.
In the vicinity of the clean fixed point representing this transition,
the disorder strength $g_M$ is marginally
irrelevant. This implies 
for the critical DoS a logarithmic correction term (see
Appendix~\ref{s-a1})
of the form
\begin{equation}
\varrho(E) = \frac{|E|}{2\pi}\left(1 + \frac{2g_M}{\pi}\ln
\frac{1}{|E|} \right).
\label{e2}
\end{equation}
%% where $\Delta$ is the ultraviolet cutoff (bandwidth). 
Clearly, since this calculation is for a system with weak, homogeneous
disorder, its relevance for behavior in the CF model with
dilute, strong scatterers needs to be tested.

In the localized phases, past work suggests several possibilities for
the behavior of the DoS near $E{=}0$. The simplest approach is to
imagine that the 
sample can be divided into independent regions of size set by the
localization length and that the contribution of each region to the
DoS can be obtained from random-matrix theory for this symmetry 
class,\cite{altland} giving finite $\varrho(E)$ at $E{=}0$.
Alternatives are suggested by the fact that the off-critical, 
disorder-free model ($p{=}0$, $\sin^2(\alpha) {\not=} 0$) has a gap in the
DoS around $E=0$. As has been well-studied in one-dimensional 
system, rare disorder configurations (termed Griffiths
strings) may fill in this gap, generating a DoS that varies as a
positive or negative power of $|E|$ near
$E{=}0$.\cite{motrunich00,motrunich01} Behavior of this kind has been
found recently in the network model representation of the
RBIM.\cite{mildenberger04}

The purpose of this article is to present numerical studies
of the behavior
of the DoS in the CF model.
An outline is as follows. 
In Sec.~\ref{s2}, we describe the model and our numerical methods. 
Section~\ref{s3} is central for the paper and contains our main findings. 
We analyze first (Sec.~\ref{s3.1}) 
the low-energy DoS in the
metallic phase (at large $p$) and confirm the logarithmic divergence, 
Eq. (\ref{e1}). As an additional manifestation of metallic behavior,
we find
random-matrix-theory  (RMT) oscillations in the DoS superseding 
this logarithmic behavior  at lowest energies. 
At smaller $p$,
one enters a localized phase (Sec.~\ref{s3.2})
where neither a logarithmic divergence nor 
oscillations are
observed. Instead,
the DoS remains finite at zero energy.  
Finally, in Sec.~\ref{s3.3}, we turn to the plateau transition
which occurs for low $p$ on the self-dual line
$\sin^2(\alpha){=}1/2$
(see Fig.~\ref{f1}). We find in a wide energy interval that the behavior
of DoS is consistent with the RG result, Eq.~(\ref{e2}). 
 In Sec.~\ref{s4} we summarize our findings and discuss directions
for future work. A brief description of an alternative microscopic model and
results obtained for it are given in Appendix \ref{s-a2}: these
results are very close to those for the CF
model, supporting the idea that the latter is generic.

%%%%%%%%%%%%%%%%%%%%%%%%%%%%%%%%%%%%%%%%%%%%%%%%%%%%%%%%%%%%%%%%
%%%%%%%%% Model
%%%%%%%%%%%%%%%%%%%%%%%%%%%%%%%%%%%%%%%%%%%%%%%%%%%%%%%%%%%%%%%%
\section{Model}
\label{s2}

\begin{figure}[tb!p]
\includegraphics[width=0.5\columnwidth]{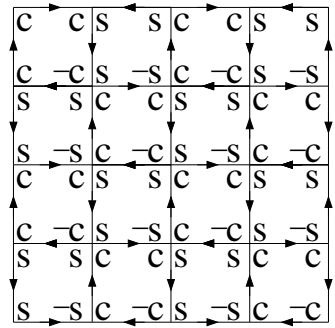}
\caption{Network representation of the clean Ising model.
The plaquettes form two sublattices, C and S. The symbols ``s'' and ``c''
 denotes the amplitudes for left and right turns, $\pm\sin\alpha$ and
 $\pm\cos\alpha$, governing the evolution at the network nodes.
 Adding disorder by
 inserting a vortex pair corresponds to flipping the signs of either a pair 
 of s or a pair of c that are associated with a given node.}
\label{f2}
\end{figure}

The CF model\cite{cho97} 
belongs to the family of network models first proposed in
Ref.~\onlinecite{chalker87} for the description of the
(conventional) quantum Hall effect.
The structure of the network in the clean limit is shown  in Fig. \ref{f2}.
The model describes quantum dynamics of noninteracting particles 
living on the directed links of a square network.
If (as here) the wave function on a link has one component,
the state of the system as a whole is represented by
an $N$-component vector, where $N$ is the number of links. 
The time evolution is
characterized by a $N\times N$ unitary 
matrix  $\mathbf U$, which in the absence of
disorder is governed by a single parameter, $0<\alpha<\pi/2$.
At every node of the
network the particle turns right or left with a probability
amplitude $\pm\cos \alpha $ or $\pm\sin \alpha$, respectively.
The amplitude signs are shown in Fig.~\ref{f2} and ensure  unitarity of the
evolution operator $\mathbf U$.\cite{klesse95} 
Note that each plaquette of the
network carries half a flux quantum so that particles pick up a
phase factor $\pi$ when moving around it.

%%%%%%%%%%%%%%%%%%%%%%%%%%%%%%%%%%%%%%%%%%%%%%%%%%%%%%%%%%%%%%%%
%%%%%%%%%%%%%%%%%%%%%%%%%%%%%%%%%%%%%%%%%%%%%%%%%%%%%%%%%%%%%%%%

In the absence of disorder, the
parameter $\alpha$ is the same for all the nodes; then the network is a
fermionic representation of the clean Ising model and  $\mathbf U$
can be diagonalized by Fourier transform.
Near the critical point and for small wave vectors $k$, the spectrum has 
a Dirac form $E = (k^2 + \xi^{-2})^{1/2}$.
The gap $\xi^{-1}= |\alpha-\pi/4|$ 
vanishes at the critical point  
$\alpha= \alpha_c \equiv \pi/4$ where the DoS is linear in energy:
$\varrho(E) = |E|/2\pi$. 

One can think of the evolution $\mathbf U$ as being generated by a
Hamiltonian. Taking ${\mathbf U} =\exp(-i\tilde{H})$, the symmetry condition
for class D, that $\tilde{H}$ is purely imaginary, implies that $\mathbf U$
is real. Hence,  within symmetry class D, disorder can be introduced into the model by
allowing for node-to-node variation of the parameter $\alpha$.
This can be done in a variety of ways. 
A disorder weak in the perturbative sense of
Refs.~\onlinecite{dotsenko83} and \onlinecite{bocquet00} can be realized
by drawing $\alpha$ for every node from a 
distribution with
a width $\delta\alpha$ that is small compared to the mean value.
In contrast, in the CF model, disorder is introduced as 
isolated defects by making the change
$\alpha\to -\alpha$ or $\alpha\to \pi-\alpha$,
for a subset of nodes randomly distributed with a concentration $p$. 
This amounts to flipping signs of either both
$\sin\alpha$  or both $\cos\alpha$ associated with such a node. 
This procedure can be viewed as the insertion of 
two additional half-flux lines into two plaquettes adjacent to the node
and belonging to the same sublattice, see Fig. \ref{f2}.
Note that the vortex pair appears with equal probability on the C- or
S-sublattice. It is this feature that distinguishes the CF
model from the
RBIM,\cite{cho97,merz01} 
which is obtained if all the additional vortices are placed
on the same sublattice.

Our numerical analysis centers on the matrix
$\mathbf U$ for a system with a torus geometry of size $L\times L$.
Since $\mathbf U$ is unitary, its $L^2$ eigenvalues  $e^{i E_j}$
lie on the unit circle, defining the energies $E_j$.
For a given realization of disorder, we compute
eigenstates in a vicinity of the value
$E{=}0$, where the special features of class D reveal themselves.
For this purpose, we use efficient
sparse matrix packages. \cite{num1,num2,num3}
The procedure is performed for an ensemble consisting typically of 
$10^4$ or $ 10^5$ disorder realizations,
and the DoS is obtained as an ensemble average.

%%%%%%%%%%%%%%%%%%%%%%%%%%%%%%%%%%%%%%%%%%%%%%%%%%%%%%%%%%%%%%%%
%%%%%%%%% Results and Discussion
%%%%%%%%%%%%%%%%%%%%%%%%%%%%%%%%%%%%%%%%%%%%%%%%%%%%%%%%%%%%%%%%
%
%%%%%%%%%%%%%%%%%%%%%%%%%%%%%%%%%%%%%%%%%%%%%%%%%%%%%%%%%%%%%%%%
\begin{figure}[tb!p]
\includegraphics[width=0.95\columnwidth,clip]{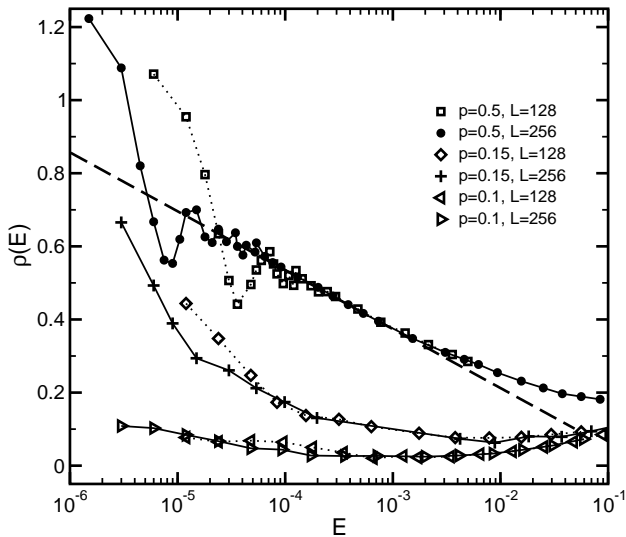}
\caption{Low-energy DoS in the metallic phase.
Parameters (upper curves): $p{=}0.5$, $\alpha{=}\pi/4$,
system sizes $L{=}128$ (squares), and $L{=}256$ (full circles).
The straight dashed line represents the logarithmic asymptotics.
For lowest energies, the RMT oscillations are clearly visible; they can
be collapsed on a single curve, as shown in Fig.~\ref{f4}. For
comparison, the results for $p=0.15$ and $p=0.1$ are also shown.}
\label{f3}
\end{figure}
%%%%%%%%%%%%%%%%%%%%%%%%%%%%%%%%%%%%%%%%%%%%%%%%%%%%%%%%%%%%%
%
\section{Results and Discussion}
\label{s3}
%%%%%%%%%%%%%%%%%%%%%%%%%%%%%%%%%%%%%%%%%%%%%%%%%%%%%%%%%%%%%%%%
%%%%%%%%% Thermal Metal
%%%%%%%%%%%%%%%%%%%%%%%%%%%%%%%%%%%%%%%%%%%%%%%%%%%%%%%%%%%%%%%%
\subsection{Thermal metal}
\label{s3.1}

Before presenting our results for the metallic phase
and comparing them to the analytical
prediction Eq.~(\ref{e1}), we briefly recall the theoretical framework
within which Eq.~(\ref{e1}) is obtained.\cite{senthil99a,bocquet00}
Using the standard procedure, one derives an effective field theory
that has the form of a diffusive supersymmetric nonlinear sigma model. 
This theory is valid on energy scales  $E < \tau^{-1}$,  where
$\tau^{-1}$ is the elastic transport scattering rate. For the lowest
energies, $E<E_{\rm Th}$, below the Thouless energy $E_{\rm Th}$
(the inverse time of diffusion through the system), the theory becomes
effectively zero-dimensional and reproduces the random-matrix theory of
class D. A renormalization-group
analysis, perturbative in the running coupling constant 
$f$ (which is proportional to $g^{-1}$),
yields
\begin{equation}
\label{e3}
  \frac{df}{d \ln \ell} = - f^2 \ ,
  \end{equation}
where $\ell$ denotes the ultraviolet cutoff. 
This implies that the infrared behavior of the system is governed by
the perfect-metal fixed point, $f\to 0$. In other words,
with increasing system size $L$, the conductance of the
metal increases logarithmically, 
$g(L) \propto \ln L$, and hence diverges in the limit
$L\to\infty$, so that the perturbative RG is
justified. A  similar analysis yields the RG equation for the second coupling
constant $\epsilon$, whose bare value is given by the energy $E$,
\begin{equation}
\label{e4}
  \frac{d\epsilon}{d \ln \ell} = (2+f)\epsilon \ ,
  \end{equation}
leading to Eq.~(\ref{e1}) for the DoS. Comparing Eqs.~(\ref{e3})
and (\ref{e4}), one sees that
the logarithmic increase in the conductance is driven by the
logarithmically divergent density of states, while
the diffusion constant $D$ remains nonsingular to this
order.  

We now turn to the results of our numerical simulations.
Fig. \ref{f3} shows the DoS calculated
at the maximal concentration $p$ of flux lines, $p{=}1/2$. The data
exhibit a logarithmic increase of the DoS over almost
three decades in $E$ for the larger system size, $L{=}256$.
We stress that the increase continues to be of logarithmic form
even though the renormalized DoS at small energies becomes
much larger than its bare (large-$E$) value $\varrho_0{\simeq} 0.1$.
This is a signature of the fact that the RG flow is toward {\it weak
coupling}, so that the one-loop result (\ref{e1}) is valid down to
arbitrarily low energies in the thermodynamic limit. 
%
%%%%%%%%%%%%%%%%%%%%%%%%%%%%%%%%%%%%%%%%%%%%%%%%%%%%%%%%%%%%%%%%
\begin{figure}[tb!p]
\includegraphics[width=0.95\columnwidth,clip]{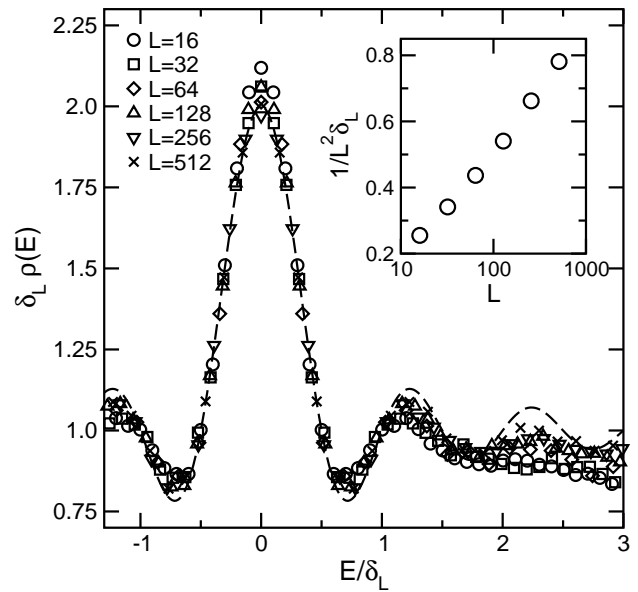}
\caption{Renormalized DoS at maximal disorder $p{=}0.5$ and
on the symmetry line $\sin^2\alpha {=}1/2$ for different
system sizes vs the energy measured in units of the level spacing $\delta_L$.
The RMT result, Eq.~(\ref{e5}), 
is plotted as a dashed line for comparison. The inset shows the 
logarithmic dependence of $1/L^2\delta_L$ on the system size $L$,
consistent with the data of Fig.~\ref{f3}. A fit of the slope
yields $\varrho_0 f_0 =0.152$ corresponding to $D=0.33$.  }
\label{f4}
\end{figure}
%%%%%%%%%%%%%%%%%%%%%%%%%%%%%%%%%%%%%%%%%%%%%%%%%%%%%%%%%%%%%%%%
%

At the smallest energies, we observe pronounced oscillations in the
DoS. These are RMT oscillations due to finite system size and serve as
another indication of the fact that we are dealing with a metallic
phase. To demonstrate the RMT origin of these oscillations, we 
replot these parts of DoS curves, rescaling the energy to the mean
level spacing $\delta_L$ at lowest energy for the corresponding system
size. Specifically, $\delta_L$ is obtained by numerically solving
$\varrho(\delta_L) = 1/L^2\delta_L$ as suggested by 
Eq.~(\ref{e5}) below.
The results are
shown in Fig.~\ref{f4} for six different system sizes.
The data collapse on a single curve, which shows that the
(renormalized) level spacing 
\begin{equation}
\label{e4a}
\delta_L=\frac{1}{L^2\varrho(E_{\rm Th})} = 
{1\over L^2\varrho_0[1+f_0\ln (L/\ell_0)]}
\end{equation}
is indeed the only relevant energy scale in the regime $E \lesssim E_{\rm
Th}$ where the RMT is applicable. 
As further shown in Fig.~\ref{f4}, the curve obtained
agrees nicely with the RMT prediction,
\begin{equation}
\label{e5}
 \varrho(E)={1\over L^2 \delta_L}\left[ 1+ {\sin(2\pi E/\delta_L)\over 
2\pi E/\delta_L}\right],
\end{equation}
 up to $E/\delta_L {\sim} 1.5$-$ 2$; for larger
energies, the oscillations are strongly suppressed. This is fully
consistent with the exponential vanishing of the RMT oscillations
beyond the Thouless energy (see, e.g., Ref.~\onlinecite{adm-physrep}).
 With increasing system size, the ratio
$E_{\rm Th}/\delta_L$ increases (though only logarithmically), so that
the RMT range includes progressively more oscillation
periods. This tendency is clearly seen in Fig.~\ref{f4}. 

For comparison, we also show in Fig.~\ref{f3} the low-energy DoS for
$p=0.15$ and $p=0.1$; the latter point is close to the expected
boundary of the metallic phase, see Fig.~\ref{f1}. It is seen
that when the system approaches the phase boundary, the
logarithmic increase of the DoS disappears and the RMT
oscillations get damped.

%%%%%%%%%%%%%%%%%%%%%%%%%%%%%%%%%%%%%%%%%%%%%%%%%%%%%%%%%%%%%%%%
%%%%%%%%% Quantum Hall Insulator
%%%%%%%%%%%%%%%%%%%%%%%%%%%%%%%%%%%%%%%%%%%%%%%%%%%%%%%%%%%%%%%%
\subsection{Localized phases}
\label{s3.2}

\begin{figure}[tb!p]
\includegraphics[width=0.95\columnwidth,clip]{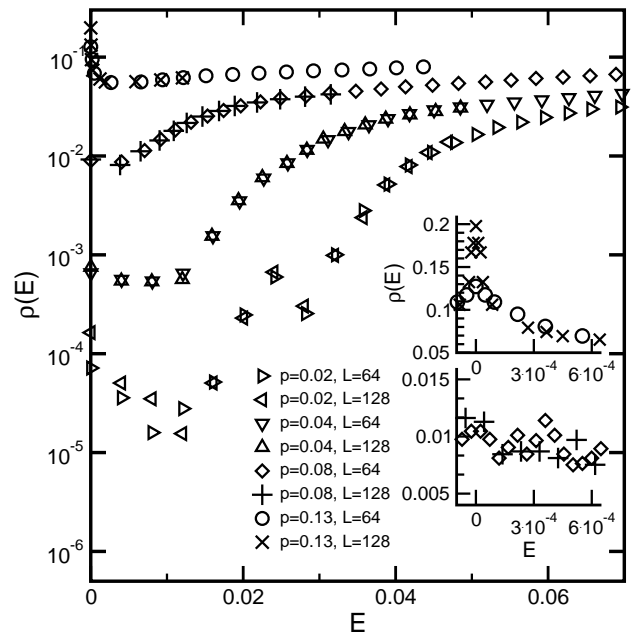}
\caption{DoS near $E=0$ for disorder values $p=0.13$, 0.08, 0.04,
  and 0.02 and for two system
sizes $L=64,\ 128$ at fixed interplaquette coupling $\sin^2(\alpha)=0.579$. 
The DoS diverges logarithmically as $E\to 0$ in the metallic phase ($p=0.13$)
and remains finite in the localized phase (other values of $p$).
The results for the lowest impurity concentration, $p=0.02$, show an
oscillatory feature induced by the band structure of the clean system, as well
strong scatter in the data at the lowest energies, which is due 
to insufficient ensemble averaging. Upper inset: Low-energy peak at
$p=0.13$; its amplitude 
increases with $L$, in agreement with  Sec.~\ref{s3.1}. Lower inset:
Low-energy DoS at $p=0.08$. No peak at $E\to0$ is detected; $\rho(E\to
0)$ is a constant independent of $L$, indicating that the system is in
the insulating phase. Statistical noise in the lower inset is more
pronounced than in the upper one due to the smallness of the DoS.}
\label{f5}
\end{figure}

Having explored the metallic phase, we now turn to the phases with
localized states.
Fixing the interplaquette coupling $\alpha$ at a value somewhat
different from $\sin^2(\alpha){=}1/2$  and decreasing the
concentration $p$ of vortex defects, we expect (see Fig.~\ref{f1})
that the system undergoes a 
transition at a critical value
$p_c(\alpha)$  (which is of the order $\sim 0.1$) and enters the
insulator or quantum Hall conductor. In both phases, bulk states are
localized.   
Figure~\ref{f5} demonstrates how this transition is reflected in the behavior
of the DoS. In this figure, we display the evolution 
of the DoS at fixed tunneling
probability $\sin^2\alpha{\simeq} 0.58$ with decreasing
defect concentration. The low-energy singular peak, whose magnitude increases
logarithmically  with the system size (as studied in Sec.~\ref{s3.1}), 
is still clearly seen for $p{=}0.13$ (see upper inset of Fig.~\ref{f5}) but
is absent once $p$ falls below 0.1  (lower 
inset of Fig.~\ref{f5}). Since this logarithmic singularity
was a key  signature of the thermal-metal phase, 
its absence signals the breakdown of the metallic
behavior and, hence, the emergence of a localized phase. 

%%%%%%%%%%%%%%%%%%%%%%%%%%%%%%%%%%%%%%%%%%%%%%%%%%%%%%%%%%%%%%%%
%%%%%%%%%%%%%%%%%%%%%%%%%%%%%%%%%%%%%%%%%%%%%%%%%%%%%%%%%%%%%%%%

For the localized phases, our main finding is that the DoS 
has a nonzero value at
$E{=}0$. It is interesting to ask how this
behavior connects with the value $\varrho(0){=}0$ expected
from Ref.~\onlinecite{bocquet00} (see Eq.~\ref{e2} above)
at the plateau transition.
Consider the DoS for $\alpha$ close to the critical value $\pi/4$, so
that the localization length $\xi$ is large. The
behavior of DoS can then be understood by using the Dirac-fermion 
RG presented in Appendix \ref{s-a1}. 
Specifically, for energies that are not too small, behavior will
be the same as at criticality, Eq.~(\ref{e2}). However, for smallest energies,
it is the localization length $\xi$ (rather than $E$) that will
terminate the RG process. In this sense, the role of $\xi$ is fully
analogous to that of finite system size $L$ at criticality. 
This implies (see Appendix~\ref{s-a1}) that $\varrho(E)$ saturates
at the value
\begin{equation}
\label{e6}
\varrho(E) \sim 
{\ell_0\over \xi}
\left(1+2{g_M\over\pi}\ln{\xi\over\ell_0}\right)^{1/2}, \qquad
  E\lesssim E_\xi,
\end{equation}
where $\ell_0$ is the ultraviolet cutoff length. The energy $E_\xi$
at which the saturation takes place is 
\begin{equation}
\label{e7}
E_\xi \sim {\ell_0\over \xi}
\left(1+2{g_M\over\pi}\ln{\xi\over\ell_0}\right)^{-1/2}.
\end{equation}
The low-energy saturation of the DoS in the localized phases which we observe
in our numerical simulations is fully consistent with these analytical
predictions. 

Before closing this subsection, we comment briefly on the regions of
localized phases where the interplaquette coupling is very weak
($\sin^2\alpha$ close to zero or to unity). As shown
recently,\cite{mildenberger04} 
in this situation the DoS of the RBIM acquires a
nonuniversal power-law singularity, $|E|^{1/z-1}$ with $z>1$
associated with Griffiths strings.\cite{motrunich00,motrunich01} 
We expect that the same mechanism should be operative in the
Cho-Fisher model as well. An analysis of these parts of the phase
diagram and of the expected Griffiths singularities is, however, outside
the scope of the present paper.  

%%%%%%%%%%%%%%%%%%%%%%%%%%%%%%%%%%%%%%%%%%%%%%%%%%%%%%%%%%%%%%%%
%%%%%%%%% quantum Hall critical line
%%%%%%%%%%%%%%%%%%%%%%%%%%%%%%%%%%%%%%%%%%%%%%%%%%%%%%%%%%%%%%%%
\subsection{Plateau transition: $\sin^2\alpha=1/2$}
\label{s3.3}

%%%%%%%%%%%%%%%%%%%%%%%%%%%%%%%%%%%%%%%%%%%%%%%%%%%%%%%%%%%%%%%%
\begin{figure}[tb!p]
\includegraphics[width=0.95\columnwidth,clip]{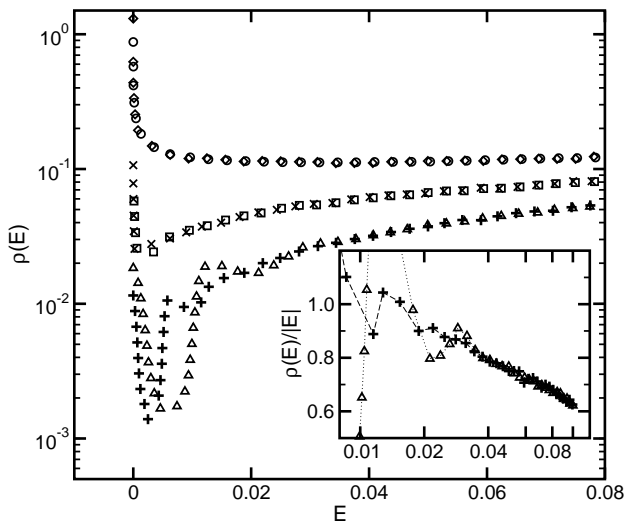}
\caption{DoS at low energy on the self-dual line $\sin^2(\alpha)=0.5$ 
for disorder concentrations
$p=0.2$ ($\circ$, $\diamond$), 0.1 ($\Box$, x), and  0.05 ($\triangle$, +),
where in each case, the first symbol is for $L=128$ and the second is
for $L=256$. 
Inset: $\varrho(E)/|E|$ at $p{=}0.05$ on a log-linear scale.
The logarithmic correction is clearly observed, in agreement with 
Eq.~(\ref{e2}).  }
\label{f6}
\end{figure}
%%%%%%%%%%%%%%%%%%%%%%%%%%%%%%%%%%%%%%%%%%%%%%%%%%%%%%%%%%%%%%%%

The phase boundary between the insulator and 
thermal Hall conductor is the location of the 
plateau transition. From the treatment of Ref.~\onlinecite{bocquet00}
(see also Appendix~\ref{s-a1}),
RG flow on this boundary is toward the clean Ising fixed point.
The corresponding RG result for the critical DoS is given by Eq.~(\ref{e2}). 
To test this prediction, we have studied the evolution of the DoS on the
self-dual line $\sin^2\alpha=1/2$ with decreasing concentration $p$ of
disorder. Results are shown in Fig.~\ref{f6}. 
For the largest two values, $p=0.2$ and $p= 0.1$, the DoS exhibits a 
peak at $E=0$ whose amplitude increases with $L$, which is a hallmark of
the metallic phase (Sec.~\ref{s3.2}). For the lowest value, $p=0.05$, this peak
is not observed anymore, suggesting that this point belongs to the phase
boundary between the two quantum Hall localized phases, see Fig.~\ref{f1}. 
Indeed, plotting $\varrho(E)/|E|$ as a function of $\log E$ (see the inset to
Fig.~\ref{f6}), we find a behavior fully consistent with the logarithmic
correction predicted by Eq.~(\ref{e2}). 
At lowest energies, an oscillatory structure is observed in the DoS curve for
$p=0.05$ in Fig.~\ref{f6}. This feature is a finite-size effect and is
governed by the few lowest-lying eigenstates which inherit information
on their position 
in the clean system. With increasing system size, the
energy window for these oscillations shrinks, so that the DoS acquires a smooth
limiting form in the thermodynamic limit.

\begin{figure}[tb!p]
\includegraphics[width=1.0\columnwidth,clip]{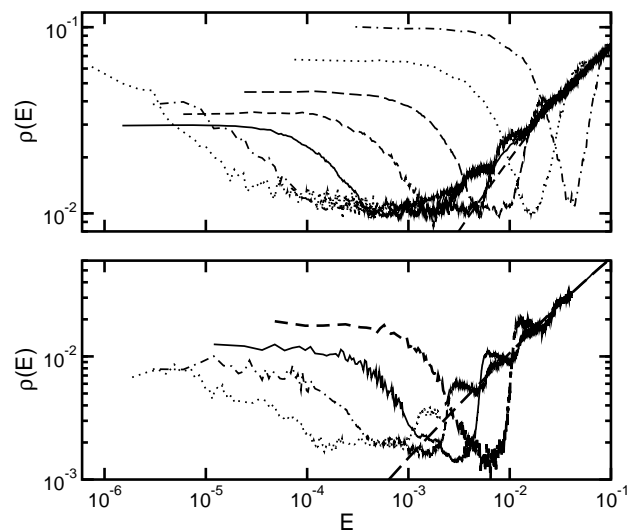}
\caption{Evolution of the DoS at low energies with increasing system size.
  DoS obtained by binning the 16 lowest-lying eigenvalues calculated
  for each network operator out of an ensemble of typically $10^4$
  disorder realizations.  
  The steplike structures in the DoS is a remnant of the
  band structure of the clean lattice and their effect gradually
  diminishes with $L$ increasing. Bold dashed lines
  show a fit to Eq. (\ref{e2}), $\varrho(E)/|E|=c_1\ln |E_0/E|$. 
  Upper panel:  $p{=}0.08$, $L{=}16,32,64,128,256,512,1024$,
  $c_1{=}0.524, E_0{=}0.433$. The position of the sharp minimum occurring at
  small system sizes, $L{=}16,32$, indicates the mean level spacing
  in these samples.  
  Lower panel: $p{=}0.05$, $L{=}128,256,512,1024$, $c_1{=}0.1875,
  E_0{=}2.865$. }
\label{f7}
\end{figure}

%%%%%%%%%%%%%%%%%%%%%%%%%%%%%%%%%%%%%%%%%%%%%%%%%%%%%%%%%%%%%%%%
We turn now to the analysis of the small-$p$  DoS at asymptotically low
energies. This requires investigation of system sizes larger than those used
in Fig.~\ref{f6}.
We show in Fig.~\ref{f7} a log-log plot of 
the DoS for the disorder concentrations 
$p=0.08$ and $p=0.05$ (i.e., below the expected position of the
three-critical fixed point, $p_T{=}0.1$, Fig. \ref{f1})
and for system sizes up to $L=1024$.
It is seen
that, when oscillations corresponding to the lowest discrete states are
discarded,  data  for different $L$ nicely combine in a single smooth
curve  corresponding to the thermodynamic-limit DoS. 
At moderately low energy,
this curve is well fitted 
Eq.~(\ref{e2}).  So, up to this point, the behavior appears to be
consistent with 
the expectation that the RG flow is directed toward the clean Ising fixed
point. However, below $E{\sim}10^{-2}$ ($E{\sim}10^{-3}$)
for $p{=}0.08$ ($p{=}0.05$), 
the DoS saturates. For $p{=}0.08$, it shows even an upturn for
$E\lesssim 10^{-3}$; curves for $p{=}0.05$ suggest a similar tendency. 

%
%%%%%%%%%%%%%%%%%%%%%%%%%%%%%%%%%%%%%%%%%%%%%%%%%%%%%%%%%%%%%%%%%%%%%
%%%%%%%%%%%%%%%%%%%%%%%%%%%%%%%%%%%%%%%%%%%%%%%%%%%%%%%%%%%%%%%%%%%%%%%%
%

We do not have an unambiguous
interpretation of these surprising findings and can only speculate about
possible  scenarios.

(i) One possibility is that the position $p_T$ of the tricritical point 
T is, in fact, not $p_T\simeq 0.1$ as was found in
Ref.~\onlinecite{chalker00}   (the phase
diagram is reproduced in our Fig.~\ref{f1}) but rather considerably smaller,
$p_T<0.05$. In addition to a conflict with the data of  
Ref.~\onlinecite{chalker00}, 
it would be quite surprising if such a numerically small value of $p_T$ should
arise. Also, it is pretty unexpected that after having reached the value of
DoS as low as 
$\varrho\sim 10^{-3}$ (lower panel of Fig.~\ref{f7}), the system 
flows toward the metallic fixed point.

(ii) A more sophisticated 
scenario that would allow reconciliation of our results with those
of  Ref.~\onlinecite{chalker00} 
is that, in fact, there are two fixed points on the
expected quantum Hall transition line $\sin^2\alpha=1/2$. Namely, in addition
to the tricritical point $p_T$, there is a repulsive fixed point at some
$p_N<p_T$. This point would then act as a ``flow splitter'' which is similar
to the role of the Nishimori point in the RBIM (hence, the subscript $N$). 

Taking this idea further, we could imagine that the tricritical point T also
has a counterpart in the RBIM, namely, the zero-temperature transition point
from the ferromagnetic phase into the spin glass. Furthermore, it is possible
that the metallic phase of CF model connects to the spin-glass line of
RBIM that exists at zero temperature for sufficiently strong disorder. 

(iii) Another nontrivial possibility is that the RG treatment of the 
theory of Dirac fermions with Gaussian random mass is, in fact, insufficient,
and %either operators related to higher moments of the distribution or
some effects -- possibly of nonperturbative origin --
eventually drive the system away from the
clean Ising fixed point. 

In order to decide which of these scenarios take place, further work
(analytical as well as numerical) is apparently needed. We will return to 
possible directions of future research in Sec.~\ref{s4}.

%%%%%%%%%%%%%%%%%%%%%%%%%%%%%%%%%%%%%%%%%%%%%%%%%%%%%%%%%%%%%%%%
%%%%%%%%% Conclusions
%%%%%%%%%%%%%%%%%%%%%%%%%%%%%%%%%%%%%%%%%%%%%%%%%%%%%%%%%%%%%%%% 
\section{Conclusions}
\label{s4}

In summary, we have presented a numerical investigation of the density of states
$\varrho(E)$ in the Cho-Fisher network model. The model 
is a generic two-dimensional representative of the symmetry class D 
describing disordered superconductors with broken spin-rotation
and time-reversal invariances, and shows the thermal quantum Hall effect.  
At a sufficiently large concentration $p$ of defects, 
the DoS  has a logarithmic divergence as
$E\to 0$ with superimposed random-matrix-theory oscillations, in agreement
with analytical predictions for the thermal-metal phase, given in
Eqs.~(\ref{e1}) and (\ref{e5}). 
Reducing $p$, we find a transition into 
localized phases (insulator and quantized Hall conductor) 
with $\varrho(E)$ finite at $E{=}0$. 

At the plateau
transition between these phases, the DoS tends to vanish as
$E\to 0$ in agreement with the behavior $\varrho(E)\sim |E|\ln(1/|E|)$
derived from 
the theory for Dirac fermions with random mass, Eq.~(\ref{e2}). 
However, at lowest $E$, this behavior breaks down, and DoS
saturates and even shows an upturn. We do not have an unambiguous explanation
for this behavior;
more work is needed in order %% to resolve this puzzle and
to understand better the properties of the system at the quantum Hall
transition line. We imagine an analysis of a phase diagram of a wider
family of class-D models, see Ref. \onlinecite{gruzberg01}.
One can also study the dependence of results on the microscopic model in
numerical simulations. In particular, a possible generalization of the
Cho-Fisher model is presented in Appendix~\ref{s-a2}. 
Further, one should study different
observables that are more susceptible to the critical behavior and 
would give additional information about the system at the
expected critical line. In particular, an analysis of the wave-function 
statistics is expected to be useful in this respect.
\cite{mildenberger05Up}

\acknowledgments We thank I.~A. Gruzberg, V.~Kagalovsky,
A.~W.~W. Ludwig, R.~Narayanan, X.~Wan, and M.~R. Zirnbauer  for fruitful
discussions. This work was supported by the SPP
``Quanten-Hall-Systeme'' and
Center for Functional Nanostructures
of the DFG and by the
A. von Humboldt Foundation.

%%%%%%%%%%%%%%%%%%%%%%%%%%%%%%%%%%%%%%%%%%%%%%%%%%%%%%%%%%%%%%%%
%% continuous distribution model
%%%%%%%%%%%%%%%%%%%%%%%%%%%%%%%%%%%%%%%%%%%%%%%%%%%%%%%%%%%%%%%%

\appendix

\section{Renormalization group for Dirac fermions with random mass}
\label{s-a1}

In this Appendix, we sketch the RG analysis of the DoS of
Dirac fermions with random mass, leading to Eq.~(\ref{e2}) at
criticality and Eq.~(\ref{e6}) away from the critical line. Our
presentation largely follows Ref.~\onlinecite{bocquet00} though we 
depart from that work at the end. The Hamiltonian has the form
\begin{equation}
\label{e-a11}
H = -i\partial_x\sigma_x-i\partial_y\sigma_y+ m({\bf r})\sigma_z,
\end{equation}
where $\sigma_\mu$ are Pauli matrices. It satisfies the Hermiticity
$H=H^\dagger$ and the class-D symmetry, $H=-\sigma_xH^T\sigma_x$. The
mass $m({\bf r})$ is a Gaussian random variable, with 
\begin{equation}
\label{e-a12}
\langle m({\bf r})m({\bf r'})\rangle = 2 g_M \delta({\bf r}-{\bf r'}).
\end{equation}
Introducing the field-theoretical representation, performing the
disorder averaging, and carrying out the RG analysis (see
Ref.~\onlinecite{bocquet00}), one gets in one-loop order the
scaling equations for renormalization of the disorder
strength $g_M$ and the energy $E$,
\begin{eqnarray}
\label{e-a13}
{d g_M\over d \ln \ell} & = & - {2 g_M^2\over \pi}; \\
{d E\over d \ln \ell} & = & \left(1+{g_M\over\pi}\right) E.
\label{e-a14}
\end{eqnarray}
Here, $\ell$ is the ultraviolet cutoff. The model is originally defined
with a microscopic cutoff $\ell_0$ (lattice spacing), so that 
$E\equiv E(\ell_0)$ and  $g_M\equiv g_M(\ell_0)$. Our aim is to
analyze the DoS, $\varrho(E)\equiv \varrho(E(\ell_0), g_M(\ell_0),
\ell_0)$. Integrating these equations, one gets
\begin{eqnarray}
\label{e-a15}
&& g_M^{-1}(\ell) =  g_M^{-1}(\ell_0) + {2\over\pi} \ln {\ell\over
  \ell_0},\\
&& E(\ell) = E(\ell_0) {\ell\over \ell_0} \left[1+ {2 g_M(\ell_0)\over\pi}
\ln {\ell\over \ell_0}\right]^{1/2},
\label{e-a16}
\end{eqnarray}
yielding 
\begin{eqnarray} 
\label{e-a17}
&& \varrho(E(\ell), g_M(\ell),\ell) = \varrho(E(\ell_0), g_M(\ell_0),
\ell_0) 
 \nonumber \\
&& \qquad  \times  {\ell\over \ell_0} \left[1+ {2 g_M(\ell_0)\over\pi}
\ln {\ell\over \ell_0}\right]^{-1/2}.
\end{eqnarray}
There are two ways in which
the renormalization can terminate:
(i) $E(\ell)$ reaches the bandwidth ($\sim 1$) 
or (ii) $\ell$ reaches the system size $L$. 
In the first case we use the fact that for large $E(\ell)$ the DoS
is essentially unaffected by disorder, $\varrho(E(\ell))\simeq
|E(\ell)|/2\pi$, yielding Eq.~(\ref{e2}) of the main text
%\begin{equation} 
%\label{e-a18}
%\varrho(E) \simeq {|E|\over 2\pi}\left[1+ 2 {g_M\over\pi}\ln {1\over
%    |E|}\right]
%\end{equation}
for $|E|\gg E_0$, 
%which is Eq.~(\ref{e2})
where $E_0$ is given by Eq.~(\ref{e-a20}). 
%
%
%%%%%%%%%%%%%%%%%%%%%%%%%%%%%%%%%%%%%%%%%%%%%%%%%%%%%%%%%%%%%%%%%%%
\begin{figure}[tb!]
%\begin{figure}[tb!p]
\includegraphics[width=0.95\columnwidth]{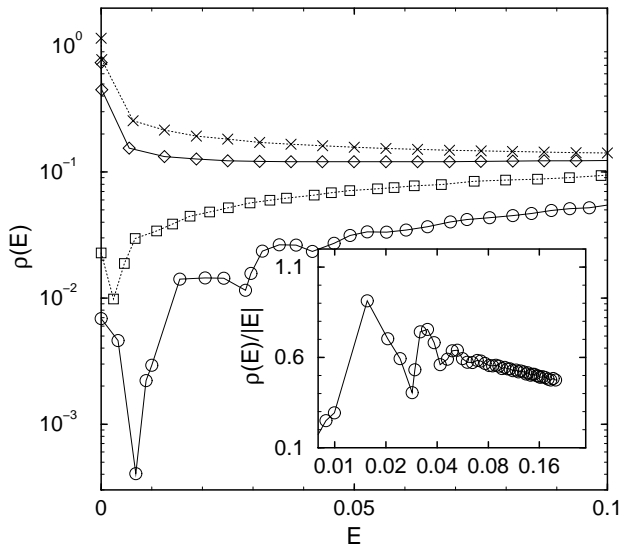}
\caption{DoS near $E=0$ on the self-dual line $\alpha_0=\pi/4$
  for the CDM with plaquette coupling parameter $\alpha$ continuously
  distributed within the window $\Delta\alpha$. 
Evolution from metallic to critical behavior is
clearly observed with decreasing disorder strength.
The four data curves (from top to bottom)are for
$\Delta\alpha=1.0$ (x),  0.8 ($\diamond$), 0.6  ($\Box$), and  0.4 ($\circ$)  
with system size $L{=}128$.
Inset: $\varrho(E)/|E|$ on the log-linear scale (see inset of
  Fig.~\ref{f6}). } 
\label{f8}
\end{figure}
%%%%%%%%%%%%%%%%%%%%%%%%%%%%%%%%%%%%%%%%%%%%%%%%%%%%%%%%%%%%%%%%%%%%%%%
%
In the second case, we get after renormalization a system with
just few degrees of freedom (ultraviolet cutoff of the order of the
system size $L$) and a small energy $E(L)\ll 1$. Since the DoS is
finite at $E{=}0$ in the random-matrix theory of class D, we find
that
\begin{equation} 
\label{e-a19}
\varrho(E) \sim {\ell_0\over L} \left[1+ 2 {g_M\over\pi}\ln
  {L\over\ell_0} \right]^{1/2}, \ \ \ E\ll E_0.
\end{equation}
Comparing Eqs.~(\ref{e2}) and (\ref{e-a20}), we find that the energy $E_0$ at which the
behavior (\ref{e2}) saturates in a finite system, crossing over into 
Eq.~(\ref{e-a19}), is
\begin{equation} 
\label{e-a20}
E_0\sim {\ell_0\over L} \left[1+ 2 {g_M\over\pi}\ln
  {L\over\ell_0} \right]^{-1/2}.
\end{equation}

Up to now, we considered a system exactly at criticality ($\alpha=\pi/4$
in the network model). Moving slightly off criticality, we enter the
localized phase, with a localization length $\xi<\infty$. Since the
states in different localization volumes are essentially independent, 
the localization length will play the same role as the
system size $L$ in Eqs.~(\ref{e2}) and (\ref{e-a20}). Substituting $\xi$
for $L$, we obtain Eqs.~(\ref{e6}) and (\ref{e7}) of the main text. 

\section{Continuous-disorder model}
\label{s-a2}

The CF model can be defined as a network model with the following
distribution of plaquette couplings:
\begin{equation}
\label{e-a1}
{\cal P}(\alpha) = (1-p)\delta(\alpha-\alpha_0) +
{p\over 2}\delta(\alpha+\alpha_0) + 
{p\over 2}\delta(\alpha+\alpha_0-\pi).
\end{equation}
It is natural to expect that properties will be
qualitatively the same for any model with a {\it generic} distribution
${\cal P}(\alpha)$. A precise
definition of the word ``generic'' is far from trivial in the present
case. In particular, we know that the distribution
\begin{equation}
\label{e-a2}
{\cal P}(\alpha) = (1-p)\delta(\alpha-\alpha_0) +
p\delta(\alpha+\alpha_0)
\end{equation}
corresponds to RBIM which does not possess a metallic phase
and thus is not generic. 

In order to test the expectation of (restricted) universality, we define
a continuous-disorder model (CDM), with a  Gaussian distribution
for the angle $\alpha$. The center $\alpha_0$
of the distribution determines the breaking of the symmetry between the
C and S plaquettes and thus governs the plateau transition. 
The width $\Delta\alpha$ determines the strength of disorder and
therefore replaces the parameter $p$ of the Cho-Fisher model.

In Fig.~\ref{f8}, we show the evolution of DoS on the self-dual line,
$\alpha_0=\pi/4$, with disorder strength decreasing from
$\Delta\alpha=1.0$ to 0.4. We see that the behavior is very similar
to that in the Cho-Fisher model, Fig.~\ref{f6}:
we observe a transition from a metal (divergent DoS) to the
critical region. Furthermore, the families of the curves look
essentially identical in both cases. We interpret this as a
confirmation of the generic character of the Cho-Fisher model. 

We have not studied the behavior of DoS at weak disorder
for lowest energies (where we observed a surprising upturn in the case
of Cho-Fisher model), relegating
a detailed investigation of this problem to future
work,\cite{mildenberger05Up} where not only spectral properties but also
those of wave functions will be analyzed.

\end{document}